\documentclass[%
reprint,
aps,
superscriptaddress,
nofootinbib,
nobibnotes,
]{revtex4-1}

\pdfoutput=1
\usepackage{graphicx}
\graphicspath{{fig/}}
\usepackage{amsmath}
\usepackage{amssymb}
\usepackage{amsfonts}
\usepackage{mathrsfs}
\usepackage{bm}
\usepackage{color}
\usepackage[%
colorlinks=true,
linkcolor=blue,
citecolor=blue,
]{hyperref}
\usepackage{tikz}
\usetikzlibrary{matrix}

\newcommand{\dif}{\mathrm{d}}

\newcommand{\br}{\bar}
\newcommand{\ck}{\check}
\newcommand{\tld}{\widetilde}
\newcommand{\ord}[2]{{}^{\mbox{\tiny$(\!#1\!)\!\!$}}{#2}}

\newcommand{\Mp}{\mathcal{M}}
\newcommand{\Mr}{\br{\Mp}}
\newcommand{\Mb}{\mathcal{M}_{0}}
\newcommand{\Sp}{\mathcal{S}}
\newcommand{\Sr}{\br{\Sp}}

\newcommand{\Vp}{\mathcal{V}}
\newcommand{\Vr}{\br{\Vp}}

\newcommand{\GW}{\mathrm{GW}}

\newcommand{\cri}{\mathrm{c}}

\newcommand{\cds}{\mathcal{D}}
\newcommand{\cdr}{\br{\nabla}}
\newcommand{\cdvr}{\br{D}}

\newcommand{\gr}{\br{g}}
\newcommand{\gb}{\ord{0}{g}}
\newcommand{\tck}{\ck{t}}
\newcommand{\uck}{\ck{u}}
\newcommand{\vck}{\ck{v}}
\newcommand{\tr}{\br{t}}
\newcommand{\ur}{\br{u}}

\newcommand{\mn}{\mu\nu}

\newcommand{\rs}{\rho\sigma}

\begin{document}

\title{Energy spectrum of gravitational waves}

\author{Rong-Gen Cai}
\affiliation{CAS Key Laboratory of Theoretical Physics, Institute of Theoretical Physics, Chinese Academy of Sciences, Beijing 100190, China}
\affiliation{School of Physical Sciences, University of Chinese Academy of Sciences, Beijing 100049, China}
\affiliation{School of Fundamental Physics and Mathematical Sciences, Hangzhou Institute for Advanced Study, University of Chinese Academy of Sciences, Hangzhou 310024, China}

\author{Xing-Yu Yang}
\email[Corresponding author.~]{yangxingyu@itp.ac.cn}
\affiliation{CAS Key Laboratory of Theoretical Physics, Institute of Theoretical Physics, Chinese Academy of Sciences, Beijing 100190, China}
\affiliation{School of Physical Sciences, University of Chinese Academy of Sciences, Beijing 100049, China}

\author{Long Zhao}
\affiliation{CAS Key Laboratory of Theoretical Physics, Institute of Theoretical Physics, Chinese Academy of Sciences, Beijing 100190, China}
\affiliation{School of Physical Sciences, University of Chinese Academy of Sciences, Beijing 100049, China}

\begin{abstract}
    The energy spectrum of gravitational waves (GWs), which depicts the energy of GWs per unit volume of space per logarithmic frequency interval normalized to the critical density of the Universe, is a widely used way for quantifying the sensitivity of GW detectors and the strength of GWs, since it has the advantage of having a clear physical meaning.
    It was found that the energy spectrum of GWs depends on the gauge when the GWs beyond the linear order perturbations are considered.
    We show that this gauge dependence issue originates from the inappropriate description for the energy of GWs.
    With the proper description for the energy of GWs, we give a well-defined energy spectrum of GWs, in which the gauge issue disappears naturally.
\end{abstract}

\maketitle

\textit{Introduction.}---%
The first direct detection of gravitational waves (GWs) has been achieved by LIGO/Virgo in 2015~\cite{Abbott:2016blz}.
Last year, NANOGrav reported a common-spectrum process which might be a stochastic gravitational wave background (SGWB)~\cite{Arzoumanian:2020vkk}.
In future, GW detectors such as SKA~\cite{Carilli:2004nx,Janssen:2014dka}, LISA~\cite{Audley:2017drz}, Taiji~\cite{Guo:2018npi}, DECIGO~\cite{Kawamura:2011zz}, and ET~\cite{Punturo:2010zz,Maggiore:2019uih}, etc., will observe more GW events in a wide range of frequencies and amplitudes from kinds of sources.
These GW observations open a new window for observing the Universe, and will shed some lights on various cosmological mysteries.
In order to calibrate the expectations for detections, it is necessary to quantify the sensitivity of instruments and the strength of their target signals.
When dealing with the sensitivity of detectors and the loudness of sources, there are three commonly used parametrizations: the characteristic strain, the power spectral density and the spectral energy density~\cite{Moore:2014lga}.
Among these parametrizations, the spectral energy density has a clear physical significance, it depicts the energy carried by GWs.
The spectral energy density is most commonly used in the studies of SGWB, and by normalizing it to the critical density of the Universe, the dimensionless energy spectrum of GWs can be constructed, which is widely used in the cosmological studies.

In the cosmological perturbation theory, due to the nonlinear coupling, the second order tensor perturbations can be induced by linear scalar perturbations.
Recently the scalar induced gravitational waves (SIGWs) have been studied widely (see~\cite{Ananda:2006af,Baumann:2007zm,Cai:2018dig,Cai:2019amo,Cai:2019elf,Cai:2019bmk,Bian:2021ini,Domenech:2021ztg}, and references therein), as they can offer the small-scale informations of the early Universe and are closely related to the primordial black holes which are attractive candidate of dark matter.
However, it was found that the energy spectrum of SIGWs depends on the gauge~\cite{Hwang:2017oxa}, which is a serious issue since the energy of GWs as a physical quantity should not depend on the gauge choice.
Actually, not only SIGWs, all GWs beyond the linear order suffer such a gauge issue.
Considering the increasing sensitivity of detectors, the computation of GWs will require going beyond the linear regime.
In cosmology, when dealing with scales much smaller than the cosmological horizon, a treatment beyond linear order might be necessary, since the linear approximation may either be not accurate enough or miss some physical effects.
In addition, it is valuable to investigate the nonlinear gravitational effects since general relativity is an intrinsically nonlinear theory.

The energy spectrum of GWs is widely used for quantifying the sensitivity of GW detectors and the strength of GWs, thus it is very important to solve this gauge issue and give a well-defined energy spectrum of GWs, otherwise it might mislead the expectations for GW detections.
In this paper, we show that in fact, the gauge dependence issue of the energy spectrum of GWs originates from the inappropriate description for the energy of GWs.
In~\cite{Cai:2021prew0071}, we gave a detailed study on the energy of GWs, and proposed a new approach to derive the energy of GWs directly from the quasilocal gravitational energy, which is valid for GWs with any wavelengths in any order of perturbations.
With this proper description for the energy of GWs, we give a well-defined energy spectrum of GWs, in which the gauge issue disappears naturally.

\textit{Energy spectrum of GWs.}---%
A widely used way for describing the strength of GWs is through their energy, which has the advantage of having a clear physical meaning.
The energy spectrum of GWs, which depicts the energy of GWs per unit volume of space per logarithmic frequency interval normalized to the critical density of the Universe $\rho_{\cri}$, is defined as~\cite{Maggiore:1999vm}
\begin{equation}
    \Omega_{\GW}(f) = \frac{\dif\rho_{\GW}/\dif \ln f}{\rho_{\cri}},
\end{equation}
in which the energy density of GWs is given by the 00-component of the Isaacson energy-momentum tensor $\rho_{\GW}=\frac{1}{4\kappa}\langle \dot{H}_{\rs} \dot{H}^{\rs} \rangle$, where $\kappa=8\pi G/c^{4}$, $H_{\rs}$ is the transverse traceless metric perturbation, and the angle brackets denote the Brill-Hartle averaging~\cite{Isaacson:1967zz,Isaacson:1968zza}.

The above definition for the energy spectrum of GWs is widely used, which works well for GWs in linear order.
However, a gauge dependence issue arises when GWs beyond the linear order are considered~\cite{Hwang:2017oxa}, which is a serious issue since the ambiguity in the theoretical predictions of GWs will mislead our expectations for their detections.
Many efforts have been devoted to this issue~\cite{Gong:2019mui,Tomikawa:2019tvi,DeLuca:2019ufz,Inomata:2019yww,Yuan:2019fwv,Giovannini:2020qta,Lu:2020diy,Ali:2020sfw,Chang:2020tji,Chang:2020iji,Chang:2020mky,Domenech:2020xin}, but there is no satisfactory solution yet.
Most discussions about this issue in literature pay attention to the concept of gauge invariance, which might miss the key for solving this issue.
Let us start with the analysis.
Firstly, why does one define the energy spectrum of GWs?
That is because one wants to describe the strength of GWs through their energy.
Then, why is the gauge dependence on the energy spectrum of GWs an issue?
That is because the energy of GWs as a physical quantity should not depend on the gauge choice.
Finally, how to describe the energy of GWs?
One may answer that the energy of GWs is described by the Isaacson energy-momentum tensor, however, this is not always the right answer, which is the very origin of the gauge issue.
The Isaacson energy-momentum tensor is valid and gauge-invariant only for linear perturbations.
Therefore, it is not surprising that there is a gauge dependence issue when one naively applies this definition to nonlinear order.
Of course one can always construct certain gauge-invariant quantity in nonlinear order~\cite{Malik:2008im,Nakamura:2020pre} and calculate the corresponding spectrum, but this spectrum might lose the physical meaning for depicting the energy of GWs.
The key for solving this gauge issue is what the proper description for the energy of GWs is.

It is well known that due to the equivalence principle the gravitational energy-momentum can not be defined locally in general relativity, hence the gravitational energy density is not a well-defined quantity, so does the energy density of GWs.
As introduced by Penrose~\cite{Penrose:1982wp}, the proper gravitational energy-momentum is quasilocal, one should measure the energy of a system by enclosing it with a closed spacelike 2-surface $\Sp$ which bounds a spacelike 3-region $\Vp$.
Since GWs are just parts of the gravitational field, their energy should also be quasilocal.
In~\cite{Cai:2021prew0071}, we gave a detailed study on the energy of GWs, and proposed a new approach to derive the energy of GWs directly from the quasilocal gravitational energy.
This proper description for the energy of GWs does not depend on the gauge and is valid for GWs with any wavelengths in any order of perturbations.
We briefly revisit our new quasilocal approach for the energy of GWs here, more details can be found in the original paper.

Basically, in order to depict the quasilocal energy of a physical system, three things are necessary: informations of the physical system, zero point of energy, and an observer.
In the physical spacetime $(\Mp,g_{\mn})$, given a spacelike 3-region $\Vp$ bounded by a closed spacelike 2-surface $\Sp$, then the desired informations of $\Vp$ are given, by such as its future-directed timelike unit normal vector field $u^{\mu}$, induced Riemannian metric $\gamma_{\mn}$ and extrinsic curvature $K_{\mn}$ with respect to $u^{\mu}$.
One needs a reference spacetime $(\Mr, \gr_{\mn})$ that corresponds to fixing the metric on the timelike boundary of the time history of the bounded region, which denotes the zero point of energy and is chosen to be Minkowski spacetime naturally.
Since the energy is a physical quantity which depends on the observer, the proper concept should be quasilocal energy of a system observed by certain observer which can be depicted by a future-directed timelike unit vector field $\tck^{\mu}$ on $\Vp$.
With some appropriate procedures one can obtain a unique embedding $\phi: \Vp \rightarrow \Mr$ which is related to $\Vp$ and $\tck^{\mu}$.
The quasilocal energy of $\Vp$ with respect to $\tck^{\mu}$ is given by the difference between the surface Hamiltonian of $\Sp$ and $\Sr \equiv \phi(\Sp)$.

Having obtained the quasilocal energy of the whole system, the energy of GWs can be derived directly, since GWs are just parts of the gravitational field in the system.
In order to identify the contributions from GWs, one can apply the scalar-vector-tensor (SVT) decomposition to the physical metric $g_{\mn}$, in which the tensor component $H_{\mn}$ depicts GWs.
The energy of GWs is the part, which is only contributed by the tensor component $H_{\mn}$, of the gravitational energy.
When doing the SVT decomposition, one needs to compare the tensors in $\Mp$ and $\Mr$, then an identification for points of these two spacetimes must be given, since it is a basic fact of differential geometry that only the comparison of tensors at the same point is meaningful.
Such an identification can be easily given by extending $\phi: \Vp \rightarrow \Mr$ into $\Phi: \Mp \rightarrow \Mr$ which shall be called the canonical reference with respect to $(\Vp,\tck^{\mu})$.

\begin{figure}[htpb]
    \centering
    \begin{tikzpicture}
        \matrix (m) [matrix of math nodes,row sep=3em,column sep=3em,minimum width=2em]
        {
            \Vp & \Mp & \color{gray}\Mb \\
            \Vr & \Mr & \color{gray}\mathbb{R}^{4}\\
        };
        \path[-stealth]
            (m-1-1) edge node[right]{$\phi$} (m-2-1)
            (m-1-2) edge node[right]{$\Phi$} (m-2-2)
            (m-1-1) -- node{$\subset$} (m-1-2)
            (m-2-1) -- node{$\subset$} (m-2-2)
            (m-1-2) edge[gray] node[below,gray]{$\Upsilon$} (m-1-3)
            (m-1-3) edge[gray] node[left,gray]{$\mathcal{Y}$} (m-2-3)
            (m-2-2) edge[gray] node[above,gray]{$\mathcal{X}$} (m-2-3);
    \end{tikzpicture}
    \caption[]{Schematic relations of physical spacetime $\Mp$, reference spacetime $\Mr$ and background spacetime $\Mb$.
        The canonical reference $\Phi$ is related to the observer, while the gague $\Upsilon$ and the coordinate $\mathcal{X}$, $\mathcal{Y}$ are arbitrary.
        The black parts are physically relevant and the grey parts are just mathematically relevant.
    }
    \label{fig:rela_st}
\end{figure}
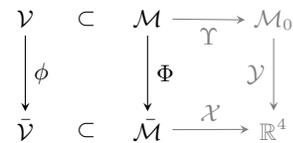

Usually in practice, one can hardly obtain the physical metric $g_{\mn}$, including GWs, by solving the Einstein equation directly.
The more feasible way is applying the perturbation theory to find the approximate solution of the Einstein equation, on a so-called background spacetime $(\Mb,\gb_{\mn})$ which is a known exact solution of the Einstein equation.
Once again, a prescription for identifying points of the physical and background spacetime is need, which is precisely the so-called gauge $\Upsilon$ in the perturbation theory.
We show the schematic relations of these spacetimes in Fig.~\ref{fig:rela_st}.

One can see that, in the physical spacetime, for given $\Vp$ with observer $\tck^{\mu}$, the total energy as well as the energy of GWs in $\Vp$ with respect to $\tck^{\mu}$ are determined.
The physical and reference spacetimes describe the dynamics and the ground state respectively, and the canonical reference is related to the observer, hence they are all physically relevant.
The background spacetime and the gauge are just mathematical auxiliary tools for solving the Einstein equation in the perturbation theory, they do not affect the physics.
In principle, one does not need perturbation theory when calculating the energy of GWs, as long as one can know the physical metric.
Therefore, it is obvious that once the proper description for energy of GWs is taken, the gauge dependence issue on the energy spectrum of GWs disappears naturally.
Recalling the physical meaning of the energy spectrum of GWs, its proper definition should be
\begin{equation}\label{eq:spct_GWs}
    \Omega_{\GW}(f;\Vp,\tck^{\mu}) = \frac{\dif[E_{\GW}(\Vp,\tck^{\mu})/\Vp]/\dif \ln f}{\rho_{\cri}}.
\end{equation}

The coordinate $\mathcal{Y}$ ($\mathcal{X}$) is a one-to-one correspondence between points in the background (reference) spacetime and points in $\mathbb{R}^{4}$, which gives a label $y^{\mu}$ ($x^{\mu}$) to points of background (reference) spacetime.
After the coordinate $\mathcal{Y}$ is given in the background spacetime, the gauge $\Upsilon$ can carry these coordinates over into the physical spacetime.
A change in the correspondence $\Upsilon \rightarrow \tld{\Upsilon}$, keeping the coordinates $\mathcal{Y}$ fixed, is called a gauge transformation, to be distinguished from a coordinate transformation $\mathcal{Y} \rightarrow \tld{\mathcal{Y}}$ which changes the labeling of points in the background and physical spacetimes together.
A gauge transformation induces a coordinate transformation in the physical spacetime, but it also changes the point in the background spacetime corresponding to a given point in the physical spacetime~\cite{Bardeen:1980kt}.
Therefore if one chooses $\{y^{\mu}\}=\{t,r,\theta,\varphi\}$ and $\Vp = \{y^{\mu}|t=T,r\le R\}$, then after the gauge transformation, the coordinates of $\Vp$ may no longer be $\{y^{\mu}|t=T,r\le R\}$.
And if one focus on $\{y^{\mu}|t=T,r\le R\}$, then after the gauge transformation, one is actually dealing with two different 3-regions in the physical spacetime.
These should be noted since sometimes confusions may arise if one neglects them.

\textit{Calculations.}---%
In the physical spacetime $(\Mp,g_{\mn})$, for a spacelike 3-region $\Vp$ bounded by a closed spacelike 2-surface $\Sp$ with observer $\tck^{\mu}$, the canonical reference $\Phi$ with respect to $(\Vp,\tck^{\mu})$ is determined, which can be obtained as follows.

For a given $\Vp$, its future-directed timelike unit normal vector field $u^{\mu}$, its induced Riemannian metric $\gamma_{\mn}$, its extrinsic curvature $K_{\mn}$ with respect to $u^{\mu}$ are all known.
Foliating $\Vp$ into a family of closed spacelike 2-surfaces $\{\Sp_{r}\}$, then for each $\Sp_{r}$ its induced Riemannian metric $\sigma_{\mn}$ with corresponding covariant derivative $\cds$ is known.
Choose the future-directed timelike unit normal vector field $\uck^{\mu}$ of $\Sp_{r}$ be same as the normal vector field $u^{\mu}$ of $\Vp$ on $\Sp_{r}$, then the outward spacelike unit normal vector field $\vck^{\mu}$ of $\Sp_{r}$ is uniquely determined and known.
Let $\tau$ be a function on $\Sp_{r}$ such that
\begin{equation}
    \cds^{\rho}\cds_{\rho}\tau / \sqrt{1+\cds^{\sigma}\tau\cds_{\sigma}\tau} = K_{\mn}\sigma^{\mn},
\end{equation}
then $\tau$ is uniquely determined and can be known by solving the equation above.
The observer $\tck^{\mu}$ on $\Sp_{r}$ is given by
\begin{equation}
    \tck^{\mu}=\sqrt{1+\cds^{\sigma}\tau\cds_{\sigma}\tau}\ \uck^{\mu} - \cds^{\mu}\tau.
\end{equation}
It is proved that there exists a unique spacelike isometric embedding $\varphi_{r}: \Sp_{r} \rightarrow \Mr$ such that on $\Sr_{r} \equiv \varphi_{r}(\Sp_{r})$ the time function (with respect to $\tr^{\mu}$) restricts to $\tau$, where $\tr^{\mu}$ is the future-directed timelike unit translational Killing vector field of the reference spacetime $(\Mr,\gr_{\mn})$.
Denote the components of the isometric embedding $\varphi_{r}$ by $\{\varphi_{r}^{\mu}\}$, which are 4 functions on $\Sp$ satisfying 4 isometric embedding equations
\begin{subequations}\label{eq:CEEs}
    \begin{align}
        \sigma_{\mn} & = (\varphi_{r}^{*}\br{\sigma})_{\mn}, \\
        \tau & = -\gr_{\mn} \tr^{\mu} \varphi_{r}^{\nu} ,
    \end{align}
\end{subequations}
where $\br{\sigma}_{\mn}$ is the induced metric of $\Sr$ in $(\Mr,\gr_{\mn})$.
Apply the aforementioned procedures to each $\Sp_{r}$ in $\{\Sp_{r}\}$, one can uniquely obtain a family of isometric embedding $\phi \equiv \{\varphi_{r}\}$ which is an embedding $\phi: \Vp \rightarrow \Mr$, and can be expanded to a mapping $\Phi: \Mp \rightarrow \Mr$ by exponential map such that  $g_{\mn}=(\Phi^{*}\gr)_{\mn}$ on $\Sp$ and $\tck^{\mu}=(\Phi^{*}\tr)^{\mu}$ on $\Vp$.
From the point of view of coordinates, the canonical reference $\Phi$ are 4 functions $\{x^{\mu}(y^{\nu})\}$ which are subject to the isometric embedding equations.
For a point in $\Mp$ labeled by $y^{\nu}$, these functions give its corresponding point labeled by $x^{\mu}$ in $\Mr$.

The procedures for obtaining the canonical reference $\Phi$ with respect to $(\Vp,\tck^{\mu})$ look a bit complicated, but they are doable and they ensure that the energy of a physical system $\Vp$ with respect to the observer $\tck^{\mu}$ is completely determined and nonnegative.
In fact, energy in general relativity is a complicated problem which has been studied widely by numerous people more than a century since general relativity was proposed by Einstein, thus one probably should not expect a quite simple procedure.

With the canonical reference $\Phi$ with respect to $(\Vp,\tck^{\mu})$, one can decompose the physical metric $g_{\mn}$ through the SVT decomposition as
\begin{equation}
    g_{\mn}=\ur_{\mu}\ur_{\nu}(\ur^{\rho}\ur^{\sigma}g_{\rs}) - 2\ur_{(\mu}(\ur^{\rho}\br{\gamma}^{\sigma}_{\nu)}g_{\rs}) + \br{\gamma}^{\rho}_{\mu}\br{\gamma}^{\sigma}_{\nu}g_{\rs},
\end{equation}
with
\begin{equation}
    \br{\gamma}^{\rho}_{\mu}\br{\gamma}^{\sigma}_{\nu}g_{\rs} = C\br{\gamma}_{\mn}+(\cdvr_{\mu}\cdvr_{\nu}-\frac{1}{3}\br{\gamma}_{\mn}\cdvr^{\rho}\cdvr_{\rho})E+\cdvr_{(\mu}F_{\nu)}+H_{\mn} ,
\end{equation}
where $\ur^{\mu}$ is the normal vector on $\Vr \equiv \phi(\Vp)$ satisfying $\uck^{\mu}=(\Phi^{*}\ur)^{\mu}$, $\br{\gamma}_{\mn} = \br{g}_{\mn} + \ur_{\mu}\ur_{\nu}$ and $\cdvr$ is its corresponding covariant derivative.
$\{C,E\}$, $F_{\mu}$ which satisfies $\cdvr^{\mu}F_{\mu}=0$, and $H_{\mn}$ which satisfies $\cdvr^{\mu} H_{\mn}=0$ and $H^{\mu}_{\mu}=0$, are called scalar-, vector-, and tensor-type components of $g_{\mn}$, respectively.
Here and hereafter the indices of these component quantities are raised and lowered with $\gr_{\mn}$.
The energy of GWs in $\Vp$ with respect to the observer $\tck^{\mu}$ is the part, which is contributed only by the tensor component $H_{\mn}$, of the gravitational energy,
\begin{equation}\label{eq:energy_GWs}
    E_{\GW}(\Vp,\tck^{\mu}) = \int_{\Vp} \mathfrak{P}_{H} \mathcal{T}^{\mu}_{\nu} \tck^{\nu} \uck_{\mu},
\end{equation}
where the operator $\mathfrak{P}_{H}$ picks out the parts of $\mathcal{T}^{\mu}_{\nu}$ that only depend on $H_{\mn}$, and
\begin{equation}\label{eq:Tmn}
    \begin{aligned}
        2\kappa\mathcal{T}^{\mu}_{\nu} = g^{\rs} [ ( \Delta^{\lambda}_{\rho\lambda}\Delta^{\mu}_{\sigma\nu} + \Delta^{\mu}_{\rs}\Delta^{\lambda}_{\lambda\nu} - 2\Delta^{\mu}_{\rho\lambda}\Delta^{\lambda}_{\sigma\nu} ) \\
        - \delta^{\mu}_{\nu} ( \Delta^{\eta}_{\rs}\Delta^{\lambda}_{\eta\lambda} - \Delta^{\eta}_{\rho\lambda}\Delta^{\lambda}_{\eta\sigma} ) ] \\
        + g^{\mu\lambda} ( \Delta^{\sigma}_{\rs}\Delta^{\rho}_{\lambda\nu} - \Delta^{\sigma}_{\lambda\sigma}\Delta^{\rho}_{\rho\nu} ),
    \end{aligned}
\end{equation}
with
$
\Delta^{\lambda}_{\mn}
=\frac{1}{2} g^{\lambda\rho} \left( \cdr_{\mu}g_{\rho\nu} + \cdr_{\nu}g_{\rho\mu} - \cdr_{\rho}g_{\mn} \right)
$
is the difference between the Christoffel symbols in $\Mp$ and $\Mr$.
$\mathfrak{P}_{H}\mathcal{T}^{\mu}_{\nu}$ can be obtained by substituting $g_{\mn}=\gr_{\mn}+H_{\mn}$ and $g^{\mn}=\gr^{\mn}-H^{\mn}+H^{\mu\rho}H^{\nu}_{\rho}-H^{\mu\rho}H^{\nu\sigma}H_{\rs}+\cdots$ into Eq.~\eqref{eq:Tmn}.
Usually in practice one does not know the physical metric $g_{\mn}$, this is the reason why the perturbation theory is needed.
One can choose any background spacetime $(\Mb,\gb_{\mn})$ and any gauge $\Upsilon$ as long as the perturbation theory is valid.
In the perturbation theory, the physical metric $g_{\mn}$ can be perturbed on the background metric $\gb_{\mn}$ as $g_{\mn}=\sum_{n=0}^{\infty} \frac{1}{n!} \ord{n}{g}_{\mn}$, then one can calculate the energy of GWs order by order.
Having obtained the energy of GWs in $\Vp$ with respect to observer $\tck^{\mu}$, the calculations for the energy spectrum of GWs is straightforward according to Eq.~\eqref{eq:spct_GWs}.

\textit{Conclusions.}---%
The energy spectrum of GWs is widely used to calibrate our expectations for GW detections, however it was found to be gauge-dependent when considering GWs beyond linear order.
Such a gauge dependence is a serious issue since it may mislead our expectations for GW detections.
In this paper, we showed that the gauge dependence issue results from the improper description for the energy of GWs.
Due to the equivalence principle the gravitational energy can not be defined locally in general relativity, hence the energy density of GWs is not well-defined.
In~\cite{Cai:2021prew0071}, we gave a detailed study on the energy of GWs, and proposed a new approach to derive the energy of GWs directly from the quasilocal gravitational energy.
With this proper description for the energy of GWs, we give a well-defined energy spectrum of GWs, in which the gauge issue disappears naturally.

We thank Misao Sasaki and Shing-Tung Yau for helpful communications.
This work is supported in part by the National Natural Science Foundation of China Grants No.11690022, No.11821505, No.11991052, No.11947302 and by the Strategic Priority Research Program of the CAS Grant No.XDPB15, and by the Key Research Program of Frontier Sciences of CAS.

\bibliography{citeLib}

\end{document}